# Phase Separation and Anomalous Volume Expansion in Frozen Microscale Eutectic Indium-Gallium upon Remelting

Se-Ho Kim,[a] Leigh T. Stephenson,[a] Alisson K. da Silva,[a] Baptiste Gault,[*a,b] Ayman A. El-Zoka[*a]

[a.] *Max-Planck-Institut für Eisenforschung GmbH, Max-Planck-Straße 1, 40237 Düsseldorf, Germany. Email: a.elzoka@mpie.de, b.gault@mpie.de*

[b.] *Department of Materials, Royal School of Mines, Imperial College, London, SW7 2AZ, United Kingdom.*

The eutectic Ga-In (EGaIn) alloy has low vapor pressure, low toxicity, high thermal and electrical conductivities, and thus has shown a great potential for smart material applications. For such applications, EGaIn is maintained above its melting point, below which it undergoes solidification and phase separation. A scientific understanding of the structural and compositional evolution during thermal cycling could help further assess the application range of low-melting-point fusible alloys. Here, we use an integrated suite of cryogenically-enabled advanced microscopy & microanalysis to better understand phase separation and (re)mixing processes in EGaIn. We reveal an overlooked thermal-stimulus-response behaviour for frozen mesoscale EGaIn at cryogenic temperatures, with a sudden volume expansion observed during in-situ heat-cycling, associated with the immiscibility between Ga and In during cooling and the formation of metastable Ga phases. These results emphasize the importance of the kinetics of rejuvenation, and open new paths for EGaIn as a self-healing material.

## I. INTRODUCTION

Most metals maintain a solid state at or near room temperature (RT) except a few: gallium (Ga), rubidium (Rb), cesium (Cs), mercury (Hg), and francium (Fr). Metals attaining a liquid state at RT bring advantages in comparison to solid metals [1]: liquid metals are for example stretchable and reformable while maintaining metal-like electrical and thermal properties. Unfortunately, rubidium's high exothermic reactivity with atmosphere, the extreme radioactivity of Cs and Fr, and the toxicity of Hg limit the potential applications of these liquid metals [2]. Gallium on the other hand, exhibits a low vapor pressure [3], low toxicity, high thermal conductivity, high electrical conductivity, and is not radioactive [4,5]. However, its melting point is 303 K, slightly higher than the commonly quoted "room temperature", which can limit its usage under ambient conditions [6].

Alloying Ga allows for decreasing its melting point, and the eutectic Ga-alloy with indium (In) is a prominent example [7]. The melting point of EGaIn is 288 K, proving advantageous for applications that rely on its electrical conductivity (3.4 x $10^6$ S $m^{-1}$) and thermal conductivity (26.43 W $m^{-1}K^{-1}$) [8,9], flexibility, deformability and ability to rejuvenate both its eutectic phase and oxide layer [4,10–14]. Thus EGaIn was found to be useful in flexible devices [15,16], cooling in micro-devices [17], self-healing reconfigurable materials [18], and actuators [19].

Although these applications have been proposed and investigated, they were tested at ambient temperature, i.e. above the melting point. Possible temperature variations were not considered and the behaviours of the material or device at temperatures below 288 K, i.e. when phase transformations would start to occur [20,21], were not considered. This raises critical questions on the operational temperature range of these devices, and on whether they could even be used in cold freezing conditions. Addressing these questions requires an understanding of how the structure and chemistry of EGaIn evolve during solidification and remelting.

Advanced microstructural investigations on EGaIn are limited, because of its temperature sensitivity. The remelting process is intuitively perceived as an uneventful transition from a solid phase with a eutectic microstructure to a liquid phase [9]. Tang et al. used in-situ transmission electron microscopy (TEM) [21] and reported on the different crystalline structures of the Ga phase achievable by controlling the cooling temperature of frozen EGaIn nanoparticles. Energy dispersive X-ray spectroscopy (EDS) was used for chemical mapping of the nanoparticles to show phase separation upon solidification, but a knowledge gap still exists regarding the qualification and quantification of the chemical and volume changes in EGaIn during cooling and remelting, especially in the more widely used microscale EGaIn droplets.

In this paper, detailed cryogenically-enabled observations reveal the intricate details of the remixing (remelting) reaction of Ga and In. We used cryo-scanning electron microscopy (SEM) combined with both a Ga- and Xe-plasma focused ion beam (FIB & pFIB respectively), cryo-transmission-Kikuchi diffraction (TKD) and atom probe tomography (APT) to capture the morphology, structure and composition of microdroplets of EGaIn (~50 μm in width on a flat Cu plate), cooled at different rates to exhibit different solidification and remixing behaviours. We report the influence of cooling and remelting on the chemistry, microstructure and volume of EGaIn microdroplets. We discuss the impact of factors such as droplet size, cooling rate and crystalline phase of Ga, and the implication these trends have on possible applications exploiting this unique feature.

## II. EXPERIMENTAL

### A. Materials & Freezing

A droplet of EGaIn was deposited on a Cu plate using the adhesive properties provided by the thin oxide surface layer (0.5 ~ 2 nm) [22], leading to the formation of stable, free-standing structures on the Cu substrate, with preserved shapes (see Figure S1 for demonstration).

Fast-cooled droplets, referred to as f-EGaIn, with dimensions of ~1 mm or less were frozen by plunging in liquid nitrogen, leading to very fast and interrupted phase separation of the liquid EGaIn into Ga- and In-rich phases (see Figure 1). Slowly-cooled droplets, referred to as s-EGaIn, were cooled on the stage of the cryo-pFIB/SEM (described below), reaching 123 K in 100 mins (estimated cooling rate of 0.02 K sec$^{-1}$).

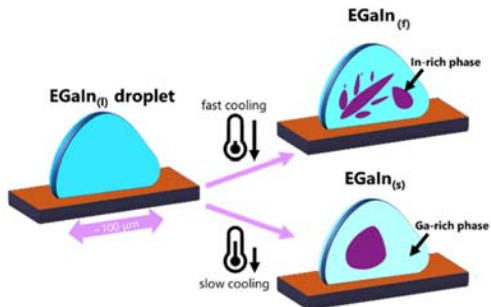

Figure 1. A schematic illustration of solidification of a liquid EGaIn micro-droplet at different cooling rates.

### B. (P)FIB/SEM Analysis

Upon freezing, samples were transferred via a cryo-enabled ultra-high-vacuum suitcase (Ferrovac GmbH) from a Sylatech glovebox into the Xe-(pFIB) (FEI Helios dual beam 600) equipped with a cryo-stage, as described in Ref. [23,24]. A Xe-pFIB is reported to cause less residual ion contamination to the specimen compared to that conventionally caused by a Ga-FIB [25]. The pFIB stage was cooled by pouring liquid nitrogen (77 K) into the attached dewar until the stage temperature reached approx. 123 K. This process took approx. 2h. The temperature profiles for the heating and cooling of the stage are shown in Figure S2. Some heating/cooling experiments (example shown in Video A) were carried out in a Ga-FIB (FEI Helios dual beam 600), only to image microdroplets that were not cross-sectioned to confirm if there are any effects induced by the pFIB milling.

We prepared atom probe specimens following the approach outlined in Ref. [26]. All ion milling and patterning steps were performed at an ion acceleration voltage of 30 kV. Both secondary- and backscattered-electron images were taken at an electron acceleration voltage and current of 30 kV and 1 nA, respectively. First, on the frozen EGaIn droplet, a ring pattern (outer/inner diameter of 300/100 μm) was set with a Xe ion beam current of 1.3 μA until a pillar with a height ~75 μm is made. To secure a clear view in atom probe measurement and to align the specimen to the atom probe local electrode, an additional rectangular trench (150 x 150 x 50 μm$^3$) was milled next to the pillar. Then, annular millings with decreasing outer and inner diameters and decreasing ion-beam current from 0.1 μA to 60 μA, 15 μA, and 1 nA were performed. Finally, a ring shape of 3/0.1 μm (outer/inner diameter) was patterned on the pillar using a beam current of 80 pA to obtain a needle-like APT specimen.

The crystallography of select APT specimens were studied at cryogenic temperatures using cryo-Transmission Kikuchi Diffraction (TKD) analysis. The patterns were collected at an acceleration voltage of 30 kV, a beam current of 2 nA, and at a stage tilt of 52°, as the specimen temperature was kept at 83 K. The acquired data was analysed using the commercial OIM software.

### C. APT on EGaIn

After fabrication, APT specimens were transferred via a cryo-UHV suitcase into a Cameca LEAP 5000 XS for atom probe analysis (see Figure 2). When operated in voltage-pulsing mode, APT analysis shows peaks pertaining to the two natural isotopes of Ga (69 and 71 Da). However, in laser-pulsing mode, despite sweeping through many experimental parameters, no peaks were detected (see Figure S3a). This may be related either to a lack of laser light absorption, or to laser-induced heating generating a liquid layer that continuously field evaporates from the apex at the DC field [27]. This is rather likely since both Ga (15 V nm$^{-1}$) and In (12 V nm$^{-1}$) have amongst the lowest evaporation fields – i.e. the critical field necessary to provoke field evaporation [28].

Analysis conditions were optimized in voltage-pulsing mode by sweeping the pulse repetition rates (Figure S3b) and stage temperature from 30, 60, to 100 K (Figure S4). Mass resolution, background level, and the homogeneity of the detector hit maps improve as the stage temperature is decreased, which indicates a high sensitivity to the stage temperature compared to other metallic samples (e.g. steel [29]). The background level at 30 K is three orders of magnitude lower than at 100 K and we were able to identify partial structural information from within the data [30]. Following parameter optimization, the specimens were analysed using a stage temperature of 30 K, voltage pulse fraction of 15%, and a pulse rate of 125 kHz. Data were reconstructed and analysed using the standard IVAS 3.8.6 visualization software (CAMECA). Details on mass spectrum analysis are discussed in the Supplementary Information (Figure S4).

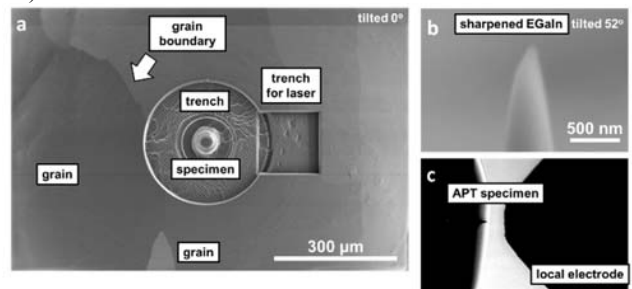

Figure 2. SEM imaging showing: (a) Top view of the final APT specimen from s-EGaIn. Note that there are grain boundaries of EGaIn. (b) An example of an APT specimen. (c) Optical image of APT specimen and local electrode inside the 5000 XS analysis chamber.

### D. In-situ APT analysis on f-EGaIn

To study the influence of temperature increase on the In content in the Ga phase, a f-EGaIn APT specimen was subjected to heat treatment by moving it from the analysis stage (30 K at $10^{-11}$ torr) into a buffer chamber at room-temperature (291 K at $10^{-9}$ torr) for increasing times (0 to 1800 sec). After each heating step, the specimen was quickly loaded onto the analysis stage again for APT measurement. This process was repeated after collected 1.5 million ions. We did not have a direct measurement of the temperature of the sample during each aging step, however, using a calculation outlined in the supplementary information, we estimated the temperature of the Cu puck after 1800 sec to be 284 K, that is lower than the melting point. The analysis was restarted right away each time at a high voltage value similar to that used before heat treatment, indicating that the specimen remained solid and its shape unmodified.

## III. RESULTS AND DISCUSSION
### A. Compositional Analysis of Metastable Ga and In Phases

Microdroplets were examined using cross-sectional secondary electron (SE) imaging in the SEM, to identify the different phases in f-EGaIn at cryogenic temperatures and how they change upon reheating (Figure 3). The phases with brighter contrast are attributed to the In-rich phase as confirmed by back-scattered electron (BSE) imaging (Figure S5), where heavier elements lead to a locally brighter contrast. In all f-EGaIn droplets, we image a dendritic structure of an In-rich phase formed in a dark Ga-rich matrix phase.

As the temperature gradually increases, from Figure 3a to 3d, we observe in-situ a complex morphological evolution of f-EGaIn during the anticipated remixing and melting. While the In-rich phase maintains its shape, the Ga-rich phase seems to melt at sub-zero temperatures (263 K), causing a volumetric expansion by a factor of approx. 2 (Figure 3b). At higher temperatures, starting from 271 K, remixing occurs with the remaining In-rich phase, which progressively disappears and is fully integrated in the liquid metal at 294 K (Figure 3d). This remelting behaviour is strikingly different from that exhibited by a s-EGaIn droplet (Figure S6), which instantaneously remelts at the eutectic temperature. Figure 4a shows a cross-sectional scanning electron micrograph of f-EGaIn microstructure, following slicing using the ion beam. As evidenced from the close up in Figure 4b, the width of the In precipitate is 2.6 ±1.7 μm with an area fraction of 6.2 % for the In-rich phase.

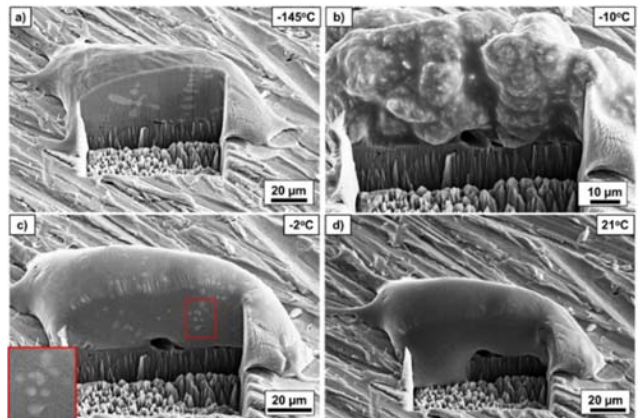

Figure 3. In-situ SEM imaging of frozen f-EGaIn droplet on a temperature-controlled stage in the pFIB whilst remelting due to gradual increase in temperature.

To explain the mechanism behind what looks like a volumetric explosion, we took a closer look at the atomic-scale chemistry in the frozen droplets. We prepared specimens for APT analysis of each of the phases, at cryogenic temperature (123 K), following the approach used previously by El-Zoka et al. [24], and guided by the BSE contrast. The successive steps leading to a suitable needle-shaped specimen for APT are depicted in Figure 4c-f. Figure 4g and 4h show the corresponding 3D atom maps of the In-rich and Ga-rich phases, respectively. Due to the rapid quenching, each phase is a mixture: The Ga-rich phase contains 1.17 ± 0.02 at. % of In that are homogeneously distributed [31] (Figure S7); the In-rich phase contains 2.74 ±0.01 at. % of Ga, which shows a slight tendency for agglomeration (Figure S7). According to the binary equilibrium phase diagram [32], slowly cooling bulk EGaIn to temperatures below the freezing point of EGaIn, i.e. 288 K, causes formation of two completely separated phases, namely orthorhombic α-Ga and fcc-In, and each of these phases do not mix with one another. A one-dimensional composition profile across the interphase interface (along the arrow) shows a rough and diffuse interface over less than 10 nm (Figure 5). Within its spatial resolution limits [33,34], APT indicates In kinetically entrapped within Ga, and vice versa, which can be ascribed to the fast cooling of the f-EGaIn sample that 'freezes' diffusing atoms in a metastable state.

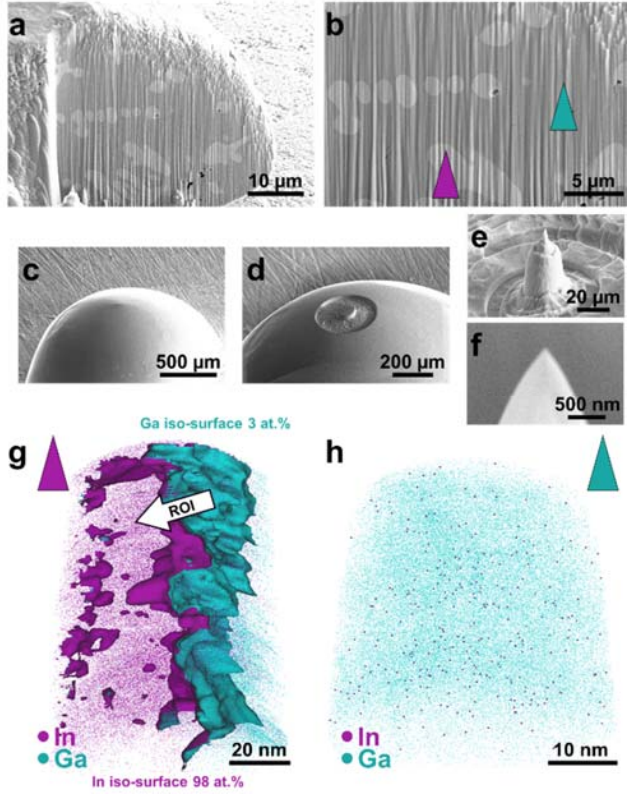

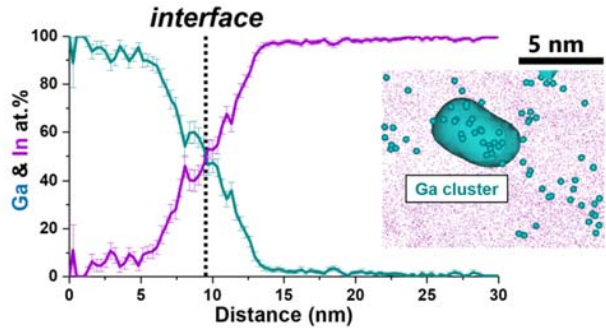

Figure 4. (a,b) Cross-sectional FIB-SEM images of f-EGaIn sample. In b, the triangles are representative positions of where APT specimens were prepared. (c-e) Cryo-specimen preparation for the liquid metal. (f) Final atom probe specimen. 3D atom maps of (g) In- and (h) Ga-rich region from f-EGaIn.

Figure 5. 1D atomic compositional profile along cylindrical region of interest across the interphase interface (φ10 x 30 nm3) in Figure 2g. Inset tomogram shows an example of locally enriched Ga atoms within the reconstructed In-rich phase in the f-EGaIn sample.

A similar set of analyses were performed on s-EGaIn droplets. The cross-sectional imaging in Figures 6a and b reveal much larger In-rich regions, 9.6 ±0.7 µm in width, almost four times larger than in f-EGaIn, with an area fraction of In-rich phase of 11.4%. APT analysis shows that only <0.0003 at.% In is detected within the Ga-rich region (see Figure 6c), supporting the extremely low solubility of In in the Ga-rich phase [35]. The In-rich phase contains 0.73 ±0.01 at.% Ga, as expected from the higher solubility of Ga in fcc-In [32]. Contrasting with f-EGaIn, the separate Ga and In phases in s-EGaIn are purer when sufficient time is given for diffusion to enable the demixing, and the compositions, reported in Table 1, confirm the low miscibility of these two metals below the eutectic point. These results confirm that the nanoscale chemistry of EGaIn will change as a function of the cooling rate, suggesting that the phase transformation occurring during solidification plays a role in the volumetric expansions observed above.

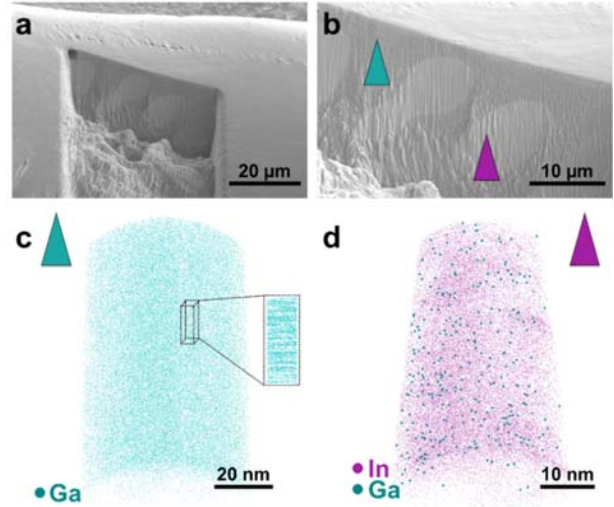

Figure 6. (a,b) Cross-sectional FIB-SEM images of s-EGaIn sample. 3D atom maps of (c) Ga region and (d) In region. The crystallography of Ga and segregation behaviour of Ga in In rich phase are discussed in Figure S8.

Table 1. Atomic compositions of each phase for f- and s-EGaIn samples obtained from APT analysis.

|  |  | Ga at.% | In at.% |
|---|---|---|---|
| f-EGaIn (plunge freezing in $N_{2(l)}$) | Ga-rich region | 98.83 ±0.02 | 1.17 ±0.02 |
|  | In-rich region | 2.74 ±0.01 | 97.26 ±0.01 |
| s-EGaIn (cooled on cryo-stage) | Ga-rich region | 99.99 - | <0.0003 - |
|  | In-rich region | 0.73 ±0.01 | 99.27 ±0.01 |

### E. In-situ Changes in the supercooled Ga-rich solid phase

As the solubility of In in the Ga-rich phase is lower than that of Ga in the In-rich phase, In diffusion is expected to be an important limiting factor in the remixing or phase growth that happens. Therefore, we performed a series of in-situ heat-treatments at room temperature, inside the atom probe ultra-high vacuum (UHV) chamber, and measured the changes in In content within the Ga-rich phase in a solid f-EGaIn droplet, to explore intraphase interactions during heating. The concentration of In increases from 1.17 to 1.74 at. %, maintaining a random distribution, after 1800 seconds of treatment as shown in Figure 7. An estimate of the sample temperature during different times of treatments can be found in Figure S9. Similar in-situ heat treatments at room temperature on a heat-sensitive Al-based alloy showed no microstructural evolution [36], which was attributed to the annihilation of non-equilibrium vacancies at the free surface of the needle-shaped specimen. However, beyond non-equilibrium vacancies from the rapid quenching of f-EGaIn, heating at or near the melting point can introduce new vacancies, thereby facilitating the diffusion of In back into the Ga-rich phase within the specimen. These measurements confirm that, at cryogenic temperatures, limited mixing of In into Ga takes place before the fusion of the Ga-rich phase.

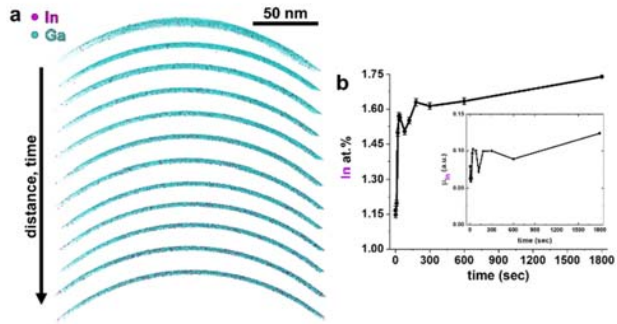

Figure 7. (a) In-situ heat treatment at room temperature of a Ga-rich APT specimen from an f-EGaIn sample for 0, 2, 10, 20, 30, 40, 80, 120, 180, 300, 600 and 1800 sec. (b) The atomic compositions of solute In atoms following each aging step. Inset shows the Pearson coefficient variation for In ($\mu_{In}$) in Ga-rich phase along with each aging time. In-In nearest-neighbor analysis indicates no segregation at all. The Pearson coefficient associated to a $\chi^2$ statistical test of the frequency distribution of In in the Ga-rich phase is measured to between 0.05 and 0.12 for all times (0 indicates complete randomness and 1 means a completely separated mixture) thereby confirming a tendency for randomness of In in Ga-rich phase.

### B. Cycling Behaviour and Origin of Volumetric Explosions

Let us now focus on the explosive volume change during remelting, to try and understand its origins and, potentially, devise ways to control it. We performed temperature cycling on an initially fast quenched EGaIn droplet, as shown in Figure 8: the same f-EGaIn droplet demonstrated explosive behaviour as expected during the first heating cycle. It was then cooled down again slowly and then re-heated for two more cycles, thereby making this droplet effectively similar to a s-EGaIn during the last two cycles. When the cooled droplet had a lower surface area of contact between the In-rich phase and the Ga-rich phase in the second and third cycles, the sudden expansion of the droplet was delayed to a higher temperature (268 K), and it did not last as long due to remixing soon after. Furthermore, we looked at the remelting of EGaIn droplets that are not cross-sectioned to eliminate any possible contribution from the Xe-plasma ion milling. A fully intact oxide layer does not inhibit the observed morphological changes, as shown in supplementary video A. These explosions were not repeatable in small droplets, with a width below 30 μm. In droplets with a width above approx. 100 μm, only limited change in volume was observed at the surface (video E). The size of the droplets reported here were expressed in terms of the shortest length, as each droplet has a non-uniform shape and thickness.

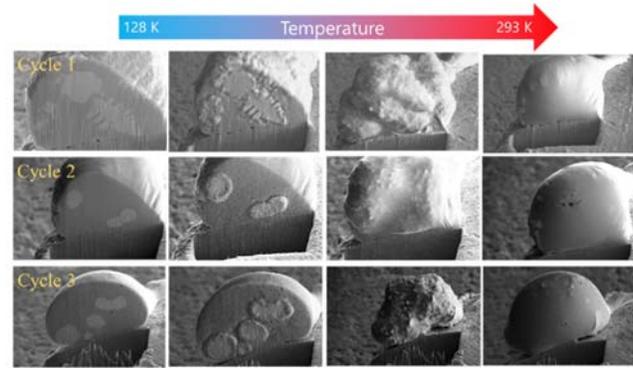

Figure 8. In-situ SEM imaging of the f-EGaIn droplet (droplet width is ~ 60 μm and field of view is ~ 90 μm x 60 μm) during several cycles melting and stage cooling. (please see supplementary videos B-D)

Based on the results shown in previous sections, we postulate that the peculiar remelting behaviour of frozen EGaIn results from the formation of metastable γ-Ga phase during the quenching at higher cooling rates resulting in higher undercooling of the liquid [37]. Ultimately, the dendritic In-rich phase in f-EGaIn in Figure 3 proves further that the cooling rate was too fast for a eutectic solidification that would typically lead to a lamellar or rod-like microstructures [38]. Thermodynamically, the smaller difference between densities of γ-Ga and liquid Ga as compared to that of α-Ga, leads to a lower nucleation barrier. Furthermore, a calculation of surface tensions between α and γ-Ga phases and In (Figure S10) shows that γ-Ga is more likely to dominate at temperatures lower than 240K.

To confirm the presence of γ-Ga in f-EGaIn due to quenching, we performed cryo-TKD directly inside the PFIB/SEM on atom probe specimens prepared from a f-EGaIn droplet following sharpening to ensure electron-transparency [39]. The indexing of the pattern gives the Cmcm space group that corresponds to the γ-Ga phase. No α-Ga phase was detected, proving the formation of γ-Ga upon fast quenching in these f-EGaIn droplets. The results are shown in Figure 9, and the pattern of Kikuchi bands can be directly attributed to γ-Ga with a high degree of confidence.

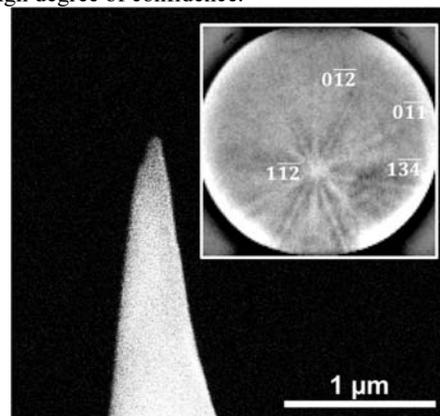

Figure 9. Frozen E-GaIn$_{(f)}$ APT specimen from cryo-sampling from Xe-FIB. Inset image shows the TKD pattern of γ-Ga on the apex.

The metastable γ-Ga shifts the entire system from eutectic to hypereutectic, as explained by the modified phase diagram in

Figure 10. Upon reheating, γ-Ga melts at a temperature lower than expected for α-Ga, in agreement with previous measurements [40,41] showing a size-effect in the melting points due to the formation of different metastable phases associated to the confinement within small Ga droplets. Tang et al. also observed by in-situ TEM study that during reheating, γ-Ga melts before In in phase-separated nanoparticles [21], and this is underpinned by a size dependence on the melting point depression in Ga-In nanoparticles, calculated by Mingear et al. using different thermodynamic models [42].

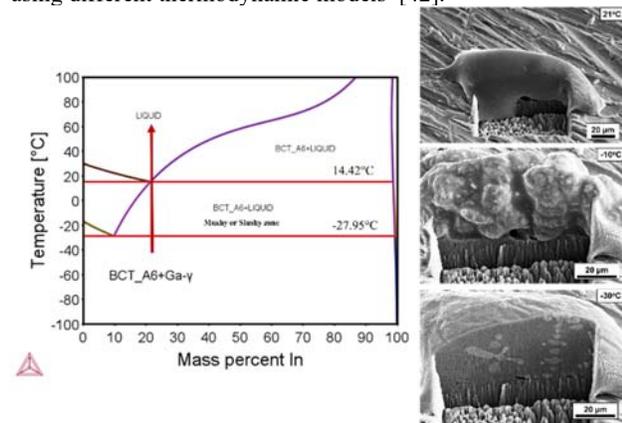

Figure 10. Explanation of the "mushy state" leading to explosion of microdroplets. The eutectic line at 14.42 °C (287.42 K) stands for the conventional α-Ga-In eutectic system. Whereas the eutectic line at -27.95 °C (245.1 K) stands for the modified system due to formation of γ-Ga as a result of undercooling at high rates.

The formation of one or more metastable Ga phases also explains the observed expansion behaviour. Metastable Ga phases have higher densities, i.e. smaller molar volumes, than α-Ga – i.e. the molar volume of γ-Ga is half of the equilibrium α-Ga [32], which has a higher molar volume than liquid Ga [7,43]. So, this expansion of frozen Ga upon heating could only be caused by a metastable γ-Ga phase rather than with α-Ga. This leads to the formation of a "mushy" state containing liquid Ga and solid In phases, which transforms into liquid with the remnant proeutectic-In phases at higher temperatures, dissolving gradually as heating increases until the eventual full remixing. This remelting behaviour was not observed in droplets with widths above approx. 100 μm (video E), due to the limited extent of undercooling in larger droplets, even when quenched in liquid nitrogen, which possibly restricts the formation of γ-Ga to areas closer to the droplet surface. Exploiting the explosive expansion of supercooled EGaIn microdroplets requires the control of droplet size and cooling rate, which dictate the crystalline structure of the Ga phase and the how finely distributed is the In-rich phase throughout the EGaIn droplet.

## IV. CONCLUSIONS

To summarize, atomic-scale compositional analyses have showed the effect of cooling rate on the phase separation that occurs in EGaIn at cryogenic temperature, highlighting the complete immiscibility of Ga and In phases, even at temperatures near the eutectic point. In-situ heating of frozen microdroplets reveals the peculiar volume expansion that occurs at sub-zero temperatures for the first time. We resort the origin of this behaviour during rejuvenation of frozen EGaIn microdroplets to expansions of metastable γ-Ga, as confirmed by cryo-TKD, phases during melting, in combination with their interaction with the In phase. We can also confirm that the higher surface area of the In phase, as defined by the extent of undercooling, the more probable is that the melting of Ga results in these explosions. We expect that future experiments could help clarify further the relationship between the surface area of In, undercooling, and size of droplet. Our results highlight the overlooked properties of EGaIn that could certainly help expand the usability of this alloy, and other low melting point alloys in general. The notion of repeatable volumetric expansion, in microscale EGaIn offer multiple opportunities for smart and flexible materials, where EGaIn droplets could be used for sensing different cooling rates and temperatures in cryogenic systems that rely on thermal cycling for functioning.

## AUTHOR CONTRIBUTIONS

A.A.Z and B.G conceptualized the investigation. S.H-K, L.T.S, and A.A.Z carried out the experimental work. A.K. carried out the thermodynamic calculations. S.H.-K and A.A.Z analysed the data. A.A. Z, B.G. and S. H-K prepared the manuscript.

## CONFLICTS OF INTEREST

There are no conflicts to declare.

## ACKNOWLEDGEMENTS

We thank Uwe Tezins, Christian Broß, and Andreas Sturm for their support to the FIB, APT, cryo-transferring suitcase, and nitrogen glovebox facilities at MPIE. S.-H.K. and B.G. also thank Dr. Andrew Breen in University of Sydney for great discussion that led us to use in-situ cryo-sample preparation for liquid sample. We also thank Dr. Jae Wung for his assistance with the cryo-TKD experiments. All authors acknowledge financial support from the ERC-CoG-SHINE-771602.

# REFERENCES

[1] T. Daeneke, K. Khoshmanesh, N. Mahmood, I. A. de Castro, D. Esrafilzadeh, S. J. Barrow, M. D. Dickey, and K. Kalantar-zadeh, *Liquid Metals: Fundamentals and Applications in Chemistry*, Chem. Soc. Rev. **47**, 4073 (2018).

[2] C. B. Eaker and M. D. Dickey, *Liquid Metal Actuation by Electrical Control of Interfacial Tension*, Appl. Phys. Rev. **3**, 31103 (2016).

[3] F. Geiger, C. A. Busse, and R. I. Loehrke, *The Vapor Pressure of Indium, Silver, Gallium, Copper, Tin, and Gold between 0.1 and 3.0 Bar*, Int. J. Thermophys. **8**, 425 (1987).

[4] G. Bo, L. Ren, X. Xu, Y. Du, and S. Dou, *Recent Progress on Liquid Metals and Their Applications*, Adv. Phys. X **3**, 1446359 (2018).

[5] H. Song, T. Kim, S. Kang, H. Jin, K. Lee, and H. J. Yoon, *Ga-Based Liquid Metal Micro/Nanoparticles: Recent Advances and Applications*, Small **16**, 1903391 (2020).

[6] K. A. Narh, V. P. Dwivedi, J. M. Grow, A. Stana, and W.-Y. Shih, *The Effect of Liquid Gallium on the Strengths of Stainless Steel and Thermoplastics*, J. Mater. Sci. **33**, 329 (1998).

[7] B. Predel and D. W. Stein, *Beitrag Zur Kenntnis Der Thermodynamischen Eigenschaften Des Systems Gallium-Indium*, J. Less Common Met. **18**, 49 (1969).

[8] D. Zrnic and D. S. Swatik, *On the Resistivity and Surface Tension of the Eutectic Alloy of Gallium and Indium*, J. Less Common Met. **18**, 67 (1969).

[9] J. N. Koster, *Directional Solidification and Melting of Eutectic GaIn*, Cryst. Res. Technol. **34**, 1129 (1999).

[10] M. D. Dickey, R. C. Chiechi, R. J. Larsen, E. A. Weiss, D. A. Weitz, and G. M. Whitesides, *Eutectic Gallium-Indium (EGaIn): A Liquid Metal Alloy for the Formation of Stable Structures in Microchannels at Room Temperature*, Adv. Funct. Mater. **18**, 1097 (2008).

[11] J.-H. So, J. Thelen, A. Qusba, G. J. Hayes, G. Lazzi, and M. D. Dickey, *Reversibly Deformable and Mechanically Tunable Fluidic Antennas*, Adv. Funct. Mater. **19**, 3632 (2009).

[12] V. T. Bharambe, J. Ma, M. D. Dickey, and J. J. Adams, *Planar, Multifunctional 3D Printed Antennas Using Liquid Metal Parasitics*, IEEE Access **7**, 134245 (2019).

[13] S. Veerapandian, W. Jang, J. B. Seol, H. Wang, M. Kong, K. Thiyagarajan, J. Kwak, G. Park, G. Lee, W. Suh, I. You, M. E. Kılıç, A. Giri, L. Beccai, A. Soon, and U. Jeong, *Hydrogen-Doped Viscoplastic Liquid Metal Microparticles for Stretchable Printed Metal Lines*, Nat. Mater. **20**, 533 (2021).

[14] D. Morales, N. A. Stoute, Z. Yu, D. E. Aspnes, and M. D. Dickey, *Liquid Gallium and the Eutectic Gallium Indium (EGaIn) Alloy: Dielectric Functions from 1.24 to 3.1 EV by Electrochemical Reduction of Surface Oxides*, Appl. Phys. Lett. **109**, 91905 (2016).

[15] L. Ren, J. Zhuang, G. Casillas, H. Feng, Y. Liu, X. Xu, Y. Liu, J. Chen, Y. Du, L. Jiang, and S. X. Dou, *Nanodroplets for Stretchable Superconducting Circuits*, Adv. Funct. Mater. **26**, 8111 (2016).

[16] M. D. Dickey, *Emerging Applications of Liquid Metals Featuring Surface Oxides*, ACS Appl. Mater. Interfaces **6**, 18369 (2014).

[17] J. Y. Zhu, S.-Y. Tang, K. Khoshmanesh, and K. Ghorbani, *An Integrated Liquid Cooling System Based on Galinstan Liquid Metal Droplets*, ACS Appl. Mater. Interfaces **8**, 2173 (2016).

[18] E. Palleau, S. Reece, S. C. Desai, M. E. Smith, and M. D. Dickey, *Self-Healing Stretchable Wires for Reconfigurable Circuit Wiring and 3D Microfluidics*, Adv. Mater. **25**, 1589 (2013).

[19] D. Kim, R. G. Pierce, R. Henderson, S. J. Doo, K. Yoo, and J.-B. Lee, *Liquid Metal Actuation-Based Reversible Frequency Tunable Monopole Antenna*, Appl. Phys. Lett. **105**, 234104 (2014).

[20] Y. Lin, Y. Liu, J. Genzer, and M. D. Dickey, *Shape-Transformable Liquid Metal Nanoparticles in Aqueous Solution*, Chem. Sci. **8**, 3832 (2017).

[21] S.-Y. Tang, D. R. G. Mitchell, Q. Zhao, D. Yuan, G. Yun, Y. Zhang, R. Qiao, Y. Lin, M. D. Dickey, and W. Li, *Phase Separation in Liquid Metal Nanoparticles*, Matter **1**, 192 (2019).

[22] K. Khoshmanesh, S.-Y. Tang, J. Y. Zhu, S. Schaefer, A. Mitchell, K. Kalantar-zadeh, and M. D. Dickey, *Liquid Metal Enabled Microfluidics*, Lab Chip **17**, 974 (2017).

[23] L. T. Stephenson, A. Szczepaniak, I. Mouton, K. A. K. Rusitzka, A. J. Breen, U. Tezins, A. Sturm, D. Vogel, Y. Chang, P. Kontis, A. Rosenthal, J. D. Shepard, U. Maier, T. F. Kelly, D. Raabe, and B. Gault, *The LaplacE Project: An Integrated Suite for Preparing and Transferring Atom Probe Samples under Cryogenic and UHV Conditions*, PLoS One **13**, e0209211 (2018).

[24] A. A. El-Zoka, S.-H. Kim, S. Deville, R. C. Newman, L. T. Stephenson, and B. Gault, *Enabling Near-Atomic–Scale Analysis of Frozen Water*, Sci. Adv. **6**, eabd6324 (2020).

[25] X. Zhong, C. A. Wade, P. J. Withers, X. Zhou, C. Cai, S. J. Haigh, and M. G. Burke, *Comparing Xe+pFIB and Ga+FIB for TEM Sample Preparation of Al Alloys: Minimising FIB-Induced Artefacts*, J. Microsc. **282**, 101 (2021).

[26] J. E. Halpin, R. W. H. Webster, H. Gardner, M. P. Moody, P. A. J. Bagot, and D. A. MacLaren, *An In-Situ Approach for Preparing Atom Probe Tomography Specimens by Xenon Plasma-Focussed Ion Beam*, Ultramicroscopy **202**, 121 (2019).

[27] S.-H. Kim, J. Y. Lee, J.-P. Ahn, and P.-P. Choi, *Fabrication of Atom Probe Tomography Specimens from Nanoparticles Using a Fusible Bi-In-Sn Alloy as an Embedding Medium*, Microsc. Microanal. **25**, 438 (2019).

[28] B. Gault, M. P. Moody, J. M. Cairney, and imon P. Ringer, *Atom Probe Microscopy*, Springer S, Vol. 160 (Springer New York, New York, NY, 2012).

[29] L. Yao, J. M. Cairney, C. Zhu, and S. P. Ringern, *Optimisation of Specimen Temperature and Pulse Fraction in Atom Probe Microscopy Experiments on a Microalloyed Steel*, Ultramicroscopy **111**, 648 (2011).

[30] B. Gault, M. P. M. P. M. P. Moody, J. M. J. M. M. Cairney, and S. P. S. P. S. P. Ringer, *Atom Probe Crystallography*, Mater. Today **15**, 378 (2012).

[31] M. P. Moody, L. T. Stephenson, A. V Ceguerra, and S. P.

Ringer, *Quantitative Binomial Distribution Analyses of Nanoscale Like-Solute Atom Clustering and Segregation in Atom Probe Tomography Data.*, Microsc. Res. Tech. **71**, 542 (2008).

[32] T. J. Anderson and I. Ansara, *The Ga-In (Gallium-Indium) System*, J. Phase Equilibria **12**, 64 (1991).

[33] B. M. Jenkins, F. Danoix, M. Gouné, P. A. J. Bagot, Z. Peng, M. P. Moody, and B. Gault, *Reflections on the Analysis of Interfaces and Grain Boundaries by Atom Probe Tomography*, Microsc. Microanal. **26**, 247 (2020).

[34] F. De Geuser and B. Gault, *Metrology of Small Particles and Solute Clusters by Atom Probe Tomography*, Acta Mater. **188**, 406 (2020).

[35] X. Y. Yin and G. S. Collins, *The Solubility of Indium in Liquid Gallium Supercooled to 12 K*, Defect Diffus. Forum **323–325**, 503 (2012).

[36] P. Dumitraschkewitz, P. J. Uggowitzer, S. S. A. Gerstl, J. F. Löffler, and S. Pogatscher, *Size-Dependent Diffusion Controls Natural Aging in Aluminium Alloys*, Nat. Commun. **10**, 4746 (2019).

[37] X. Meng Chen, G. Tao Fei, X. Feng Li, K. Zheng, and L. De Zhang, *Metastable Phases in GalliumIndium Eutectic Alloy Particles and Their Size Dependence*, J. Phys. Chem. Solids **71**, 918 (2010).

[38] G. Kaptay, *A Method to Estimate Interfacial Energy between Eutectic Solid Phases from the Results of Eutectic Solidification Experiments*, in *Materials Science Forum*, Vols. 790–791 (Trans Tech Publications Ltd, 2014), pp. 133–139.

[39] K. Babinsky, R. De Kloe, H. Clemens, and S. Primig, *A Novel Approach for Site-Specific Atom Probe Specimen Preparation by Focused Ion Beam and Transmission Electron Backscatter Diffraction.*, Ultramicroscopy **144**, 9 (2014).

[40] A. Di Cicco, *Phase Transitions in Confined Gallium Droplets*, Phys. Rev. Lett. **81**, 2942 (1998).

[41] A. Di Cicco, S. Fusari, and S. Stizza, *Phase Transitions and Undercooling in Confined Gallium*, Philos. Mag. B Phys. Condens. Matter; Stat. Mech. Electron. Opt. Magn. Prop. **79**, 2113 (1999).

[42] J. Mingear, Z. Farrell, D. Hartl, and C. Tabor, *Gallium-Indium Nanoparticles as Phase Change Material Additives for Tunable Thermal Fluids*, Nanoscale **13**, 730 (2021).

[43] B. Predel and A. Emam, *Die Volumenänderung Bei Der Bildung Flüssiger Legierungen Der Systeme Ga-In, Ga-Sn, In-Bi, In-Pb, In-Sn Und In-Tl*, J. Less-Common Met. **18**, 385 (1969).

**Supplementary Information**

**Phase Separation and Anomalous Volume Expansion in Frozen Microscale Eutectic Indium-Gallium upon Remelting**

*Se-Ho Kim, Leigh T. Stephenson, Alisson K. da Silva, Baptiste Gault\*, Ayman A. El-Zoka\**

\*co-corresponding authors

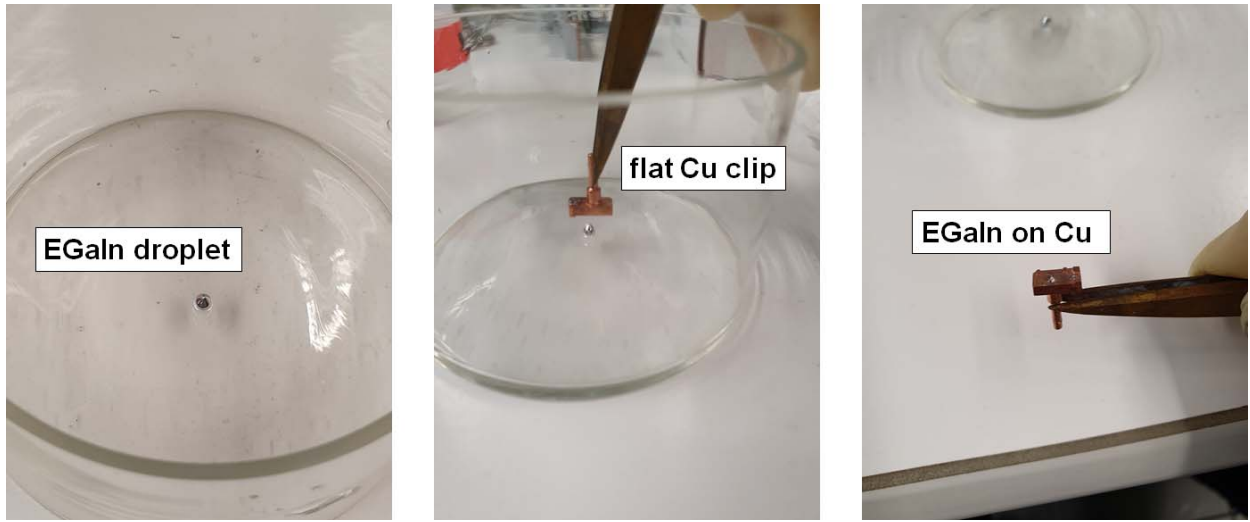

**Fig. S1.** Liquid metal sample preparation for APT specimen fabrication. First, an EGaIn droplet was placed on a clean glassware. Then, a flat Cu clip was pressed down on the droplet. Ga-oxide layer has a high adhesive strength and therefore parts of a liquid EGaIn stuck with a commercial Cu clip and it could easily be placed on the clip.

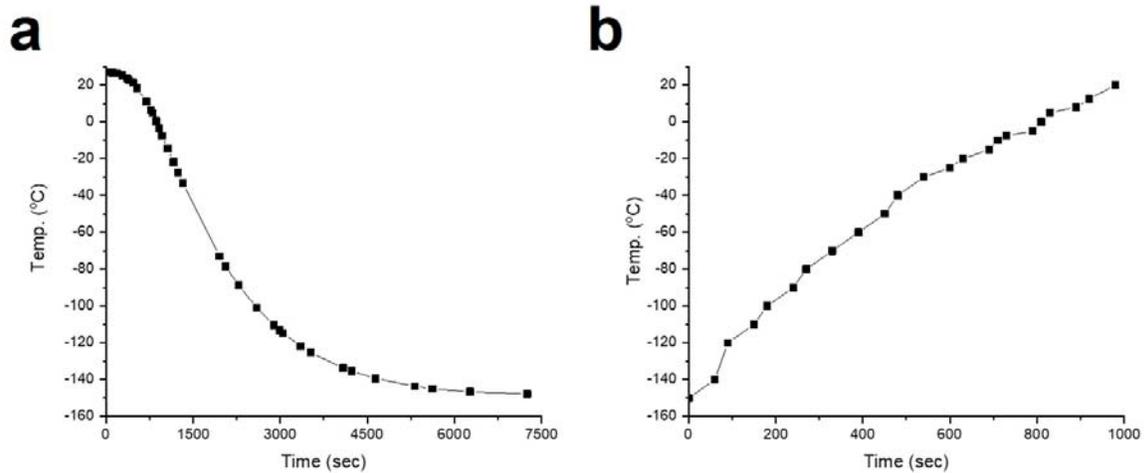

**Fig. S2.** Temperature profiles when the cryo-FIB stage is (a) cooled down to -150 °C (123 K) (with liquid nitrogen) and (b) heated up to 20 °C (293 K) (with a 40 °C (313 K) heating system).

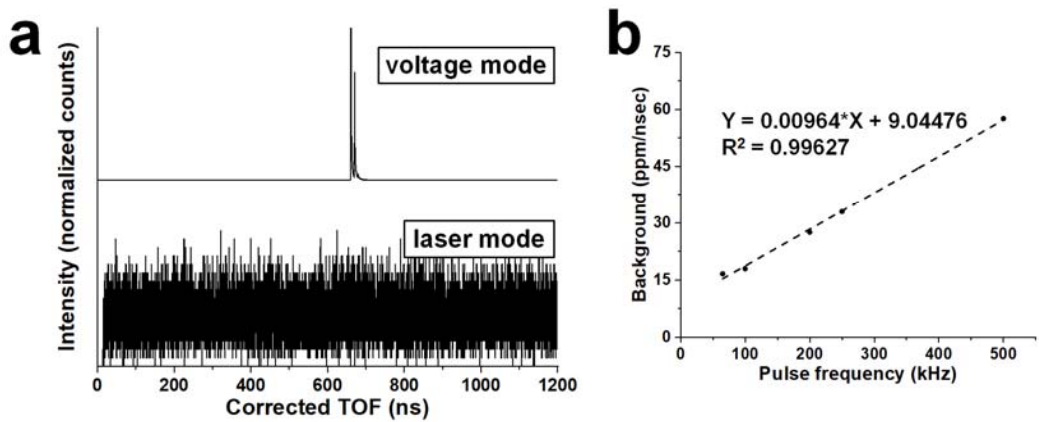

**Fig. S3.** (a) Collected time-of-flight signals from pulsed-laser and pulsed-voltage mode. (b) Pulse frequency versus background level in voltage mode. For preliminary studies, we measured the liquid EGaIn metal using both laser and voltage mode in 5000 XS system. Using the voltage mode, we collected clear signals with distinct time-of-flights, however, using laser mode we were not able to collect any noticeable signals. Therefore, we measured all liquid metal sample using voltage mode.

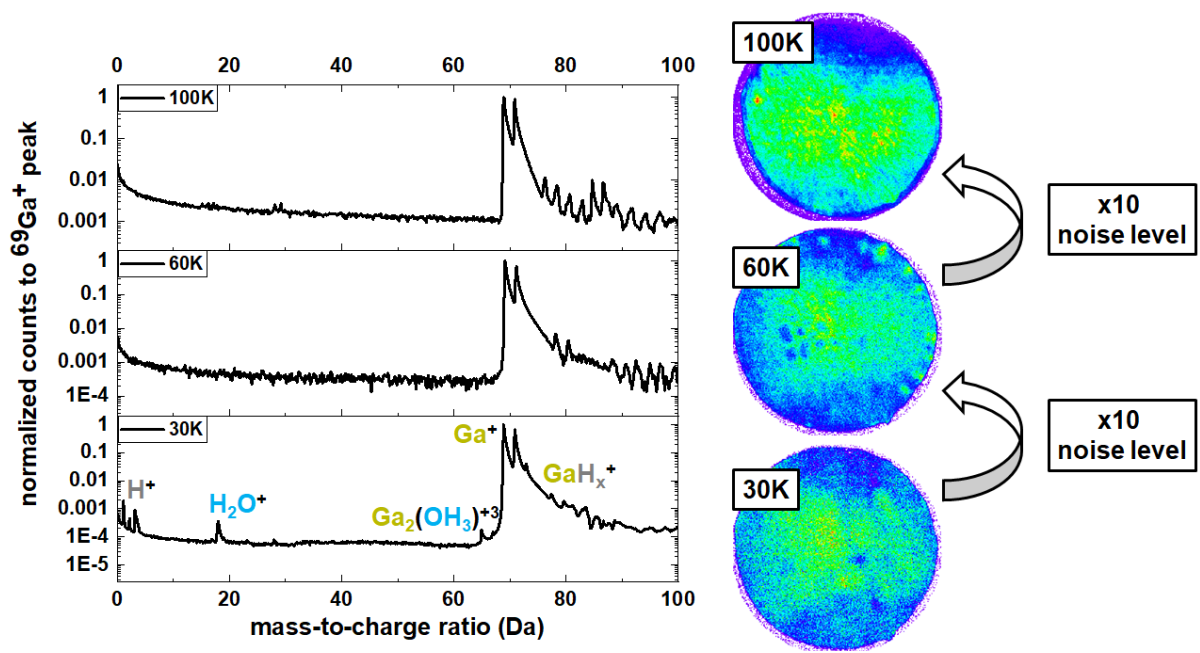

**Fig. S4.** Mass spectra and detector histograms comparison at different stage temperatures during APT analysis (30, 60, 100 K).

The mass spectrum of the corresponding 3D atom map in Fig. 6c shows that major peaks detected are $^{69}Ga^+$ and $^{71}Ga^+$. Some Ga complex peaks are also detected, as Ga evidently tends to evaporate in clusters. D.L. Barr investigated Ga ion emission from liquid Ga sources using a time-of-flight spectrometer detecting $Ga_n^+$ molecular ions with n ranging up to 30 [1].

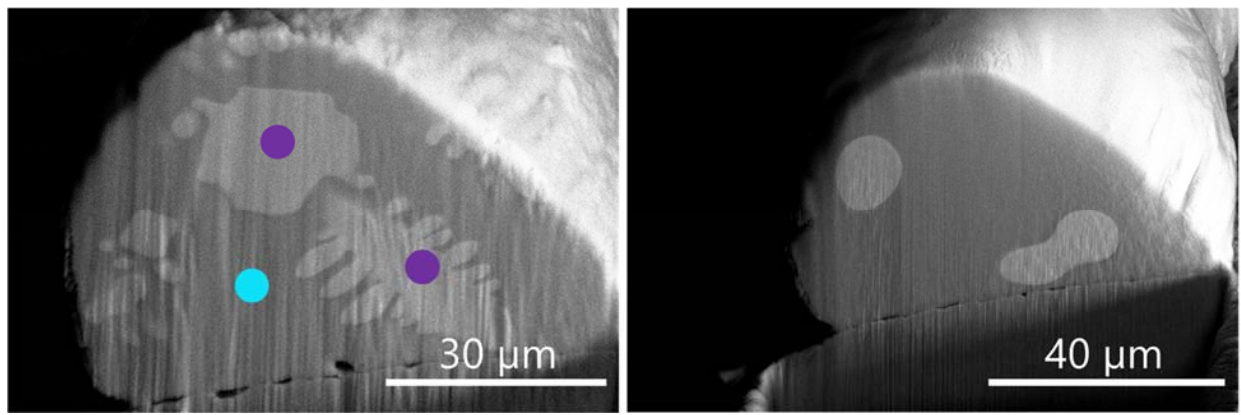

**Fig S5.** BSE Images of (a) f-EGaIn and (b) s-EGaIn droplets before remelting for allocating of In (purple dot) and Ga (light-blue dot) Phases. Bright contrast indicates a heavier element region (In).

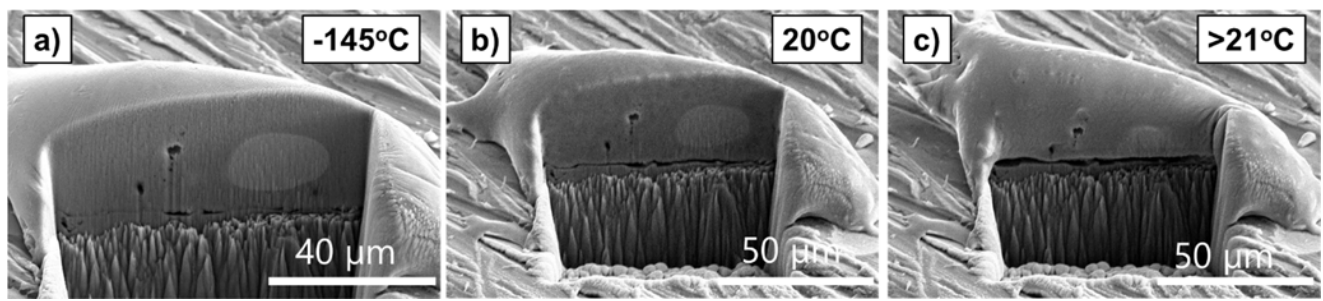

**Fig S6.** In-situ SEM imaging of the s-EGaIn droplet on a temperature-controlled stage whilst remelting due to gradual increase in temperature.

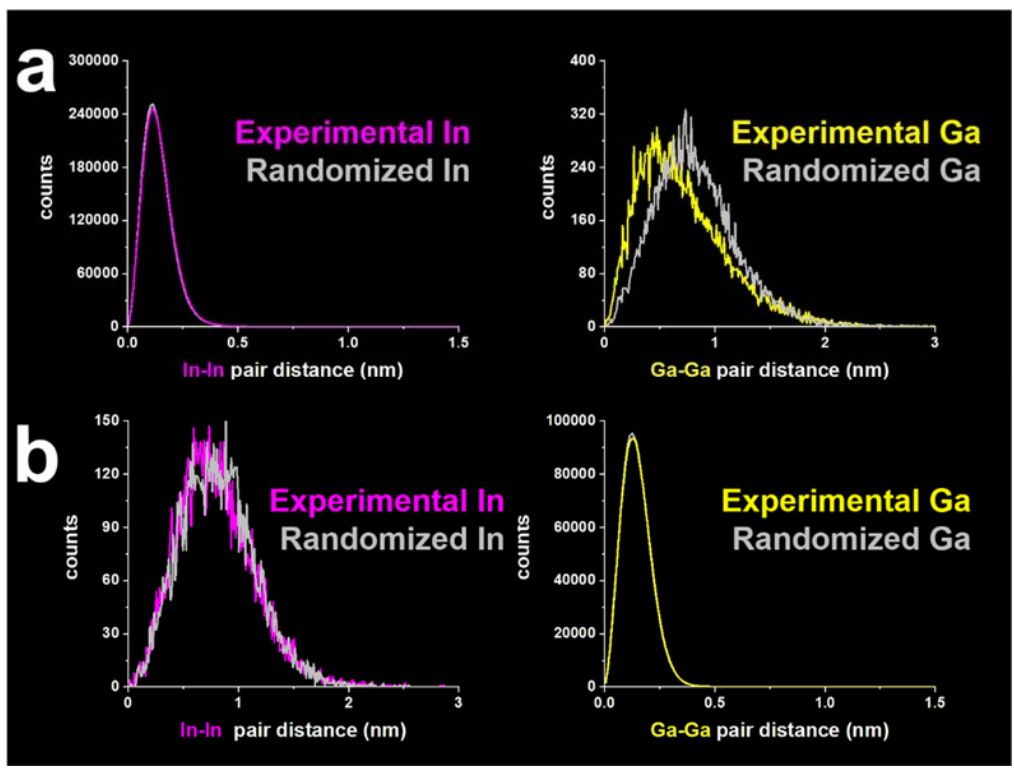

**Fig. S7.** Experimental In-In (purple line) and Ga-Ga (yellow line) first nearest-neighbor distance along with simulated Gaussian curves (grey lines) from (a) In-rich and (b) Ga-rich regions in f-EGaIn.

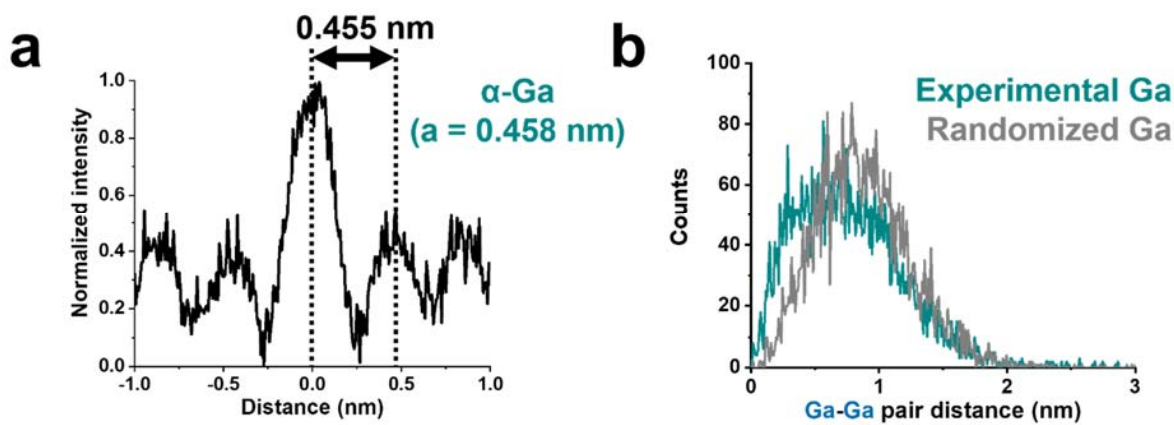

**Fig. S8.** (a) Normalized spatial distribution map of Ga atoms in Fig. 6c. (b) Ga-Ga nearest neighbor distance in Fig. 6d. We performed Ga-Ga nearest neighbor analysis from an In phase dataset. Ga phase in s-EGaIn sample was not investigated since the composition of solute In is too low to be analyzed (<10 ppm). The collected Ga-Ga distance distribution was compared with a randomized Ga distribution, in which the atomic position is unchanged but the mass-to-charge ratio are randomly swapped.

The analysis of the Ga-rich phase shows a crystallographic pole and the associated set of atomic planes in Fig 6c. The element-specific spatial distribution map (SDM) calculated along the pole orientation within the reconstructed dataset aids in determining the interplanar distance and could be used to optimize the reconstruction [2]. In this case, due to the complexity of the α, γ-Ga structures, the orientation could not yet be identified. Nevertheless, we detected a periodicity of Ga atoms (see Fig. S8a) and with aids of cryo-TKD in future work, the orientation could be identified. Fig. S4b shows the experimental Ga-Ga distribution, which deviates from the randomized Ga distribution, evidencing a tendency for Ga atoms to phase separate from the In-rich phase in the s-EGaIn sample.

*Calculating the Cu puck temperature.*

To estimate the temperature profile of EGaIn APT sample during the in-situ heat-treatment experiments, a semi-infinite solid, unsteady-state solution is used. The heat convection to the EGaIn specimen is negligible as the buffer pressure, where the sample is placed, is as low as $10^{-9}$ torr. When the Cu puck (with EGaIn specimen) is taken out from the analysis chamber to the buffer chamber, the puck is contacted with a room-temperature buffer carousel. Since the heat reservoir of a buffer carousel is much larger than that of the Cu holder, the following equation is used for the temperature estimation:

$$\frac{T(x,t) - T_s}{T_i - T_s} = \text{erf}(\frac{x}{2\sqrt{\alpha t}})$$

, where $T_s$ is the resevoir's surface temperature (i.e. buffer carousel temperature), $T_i$ is the initial temperature of Cu puck, $x$ is distance from the contacted surface to EGaIn sample (0.02m), $t$ is contacted time in sec, and $\alpha$ is the thermal diffusivity of Cu (0.000111 m² sec⁻¹).

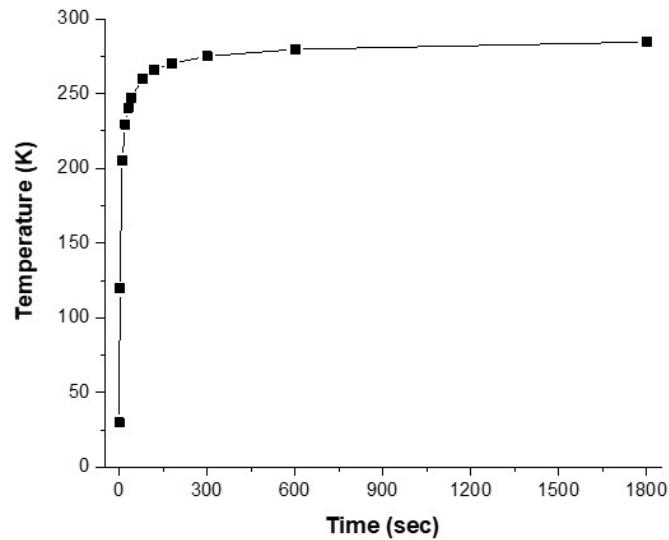

**Fig. S9.** Temperature-time profile for the surface of the Cu clip at different time contacted with a buffer carousel.

The calculated temperature profile of the Cu clip is shown in Fig. S9. Although the EGaIn sample was taken out for 1800 sec on a buffer chamber, it did not undergo a phase transformation to a liquid phase and the APT was performed without any noticeable micro-fractures. The estimated temperature of a Cu holder is calculated to be 11.4 °C (284.4 K) which is still below the melting temperature of EGaIn. Moreover, considering the additional heat transfer to a cooled bulk EGaIn droplet from a heated Cu clip, a lower temperature is likely expected at the maximum time of 1800 sec. While we saw phase changes and explosion in micro droplets at temperatures close to 0 °C (273 K). These in situ APT experiments could only be done on much bigger droplets where no explosion was observed. Thus, the rise in temperature of the droplet above 0 °C (273 K) did not change the shape of the tip.

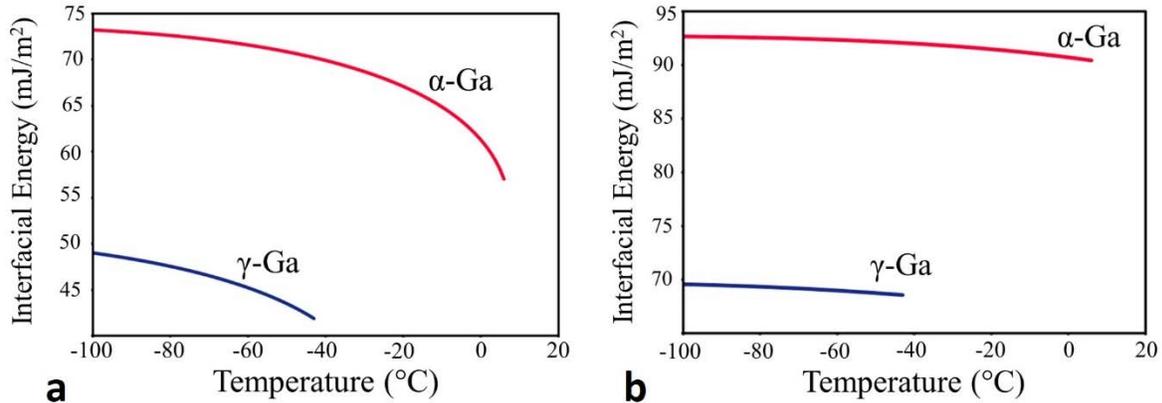

**Fig S10.** Coherent interfacial energy between the possible Ga-rich phases (α-Ga and γ-Ga) and (a) the liquid phase and (b) BCT-A6 In-rich phase. The values were calculated using the Becker's bond energy approach as implemented in the Thermo-Calc software in conjunction with the G35 Binary semiconductors thermodynamic database. Values should be used with care, since interfacial energy is dependent on many factors that are ignored in the coherent energy estimations (e.g., incoherency, orientation, and curvature).[3]

**References**

1   D. L. Barr, *J. Vac. Sci. Technol. B Microelectron. Process. Phenom.*, 1987, **5**, 184–189.

2   M. P. Moody, B. Gault, L. T. Stephenson, D. Haley and S. P. Ringer, *Ultramicroscopy*, 2009, **109**, 815–824.

3   R. Becker, *Ann. Phys.*, 1938, **424**, 128–140.